\documentclass{article}

\usepackage[utf8]{inputenc} 
\usepackage[T1]{fontenc}    
\usepackage{hyperref}       
\usepackage{url}            
\usepackage{booktabs}       
\usepackage{amsfonts}       
\usepackage{nicefrac}       
\usepackage{microtype}      
\usepackage{lipsum}
\usepackage{graphicx}
\usepackage[rightcaption]{sidecap}
\usepackage{float}
\usepackage[all]{nowidow}

\begin{document}

\title{Band power modulation through intracranial EEG stimulation and its cross-session consistency}

\author{\\Christoforos A Papasavvas$^{1}$, Gabrielle M Schroeder$^{1}$, Beate Diehl$^{3}$,\\Gerold Baier$^{4}$, Peter N Taylor$^{1,2,3}$, Yujiang Wang$^{*1,2,3}$}

\maketitle 

\begin{enumerate}
\item{CNNP Lab (www.cnnp-lab.com), Interdisciplinary Computing and Complex BioSystems Group, School of Computing, Newcastle University, Newcastle upon Tyne, United Kingdom}
\item{Faculty of Medical Sciences, Newcastle University, Newcastle upon Tyne, United Kingdom}
\item{UCL Queen Square Institute of Neurology, Queen Square, London, United Kingdom}
\item{Cell \& Developmental Biology, Division of Biosciences, University College London, London, United Kingdom}
\end{enumerate}
* Yujiang.Wang@newcastle.ac.uk
\newpage
\begin{abstract}





Background: Direct electrical stimulation of the brain through intracranial electrodes is currently used to probe the epileptic brain as part of pre-surgical evaluation, and it is also being considered for therapeutic treatments through neuromodulation. It is still unknown, however, how consistent intracranial direct electrical stimulation responses are across sessions, to allow effective neuromodulation design.  
 
Objective: To investigate the cross-session consistency of the electrophysiological effect of electrical stimulation delivered through intracranial EEG. 
 
Methods: We analysed data from 79 epilepsy patients implanted with intracranial EEG who underwent brain stimulation as part of a memory experiment. We quantified the effect of stimulation in terms of band power modulation and compared this effect from session to session. As a reference, we applied the same measures during baseline periods. 
 
Results: In most sessions, the effect of stimulation on band power could not be distinguished from baseline fluctuations of band power. Stimulation effect was also not consistent across sessions; only a third of the session pairs had a higher consistency than the baseline standards. Cross-session consistency is mainly associated with the strength of positive stimulation effects, and it also tends to be higher when the baseline conditions are more similar between sessions.
 
Conclusion: These findings can inform our practices for designing neuromodulation with greater efficacy when using direct electrical brain stimulation as a therapeutic treatment.

\end{abstract}


\section{Introduction}

About 35\% of patients with epilepsy are drug-resistant and require additional treatment \cite{Schmidt2009, deTisi2011}. In this context, direct electrical stimulation through intracranial electroencephalography (iEEG) has become an invaluable tool for clinicians. Direct electrical stimulation is currently used in three ways. First, functional mapping of the cortex so that eloquent cortical areas are preserved in resective epilepsy surgery \cite{Lesser1994, Gollwitzer2018}. Second, measuring the “epileptogenicity” of the stimulated and surrounding areas \cite{Valentin2005}. Third, exploring the neuromodulatory potential of direct electrical stimulation which can be the basis for therapeutic interventions \cite{Keller2018, Khambhati2019}. In this work we will focus on the neuromodulatory potential of intracranial electric stimulation. Arguably, to achieve any therapeutic goals, the effect of stimulation should be consistent across multiple sessions \cite{Guerra2020}. To our knowledge, the consistency of iEEG stimulation effect has not been studied systematically.

Neuromodulation has been explored as an alternative treatment for patients with non-conclusive pre-surgical evaluation of the epileptogenic zone \cite{Boon2009}. In such cases, without any candidate resection area, the goal is to modulate the epileptic network in a way that enhances physiological neural activity, and prevents pathological, or seizure activity. It is currently unknown how a targeted modulatory effect can be achieved \textit{a priori}, but several studies have begun to map out how stimulation affects the brain both electrophysiologically as well as behaviourally. For instance, Keller \textit{et al.} showed that repeated stimulation modulated the excitability of neighbouring areas around the stimulation site \cite{Keller2018}. Memory enhancement has been reported after using a closed-loop electrical stimulation of the lateral temporal cortex \cite{Ezzyat2018}. Furthermore, stimulation applied to the posterior cingulate cortex induced an increase of low gamma power in hippocampus which correlated with the magnitude of memory impairment \cite{Natu2019}. Muller and colleagues have reported a correlation between the modulation of high gamma frequencies and somatosensory perception, both induced by direct current stimulation \cite{Muller2018}. Khambhati and colleagues demonstrated functional reconfiguration of brain networks after stimulation as indicated by alterations in band-specific functional connectivity \cite{Khambhati2019}, while Huang and colleagues further demonstrated the close relationship of functional connectivity and stimulation-induced band power modulation \cite{Huang2019}. Similarly, another study showed that temporal cortex stimulation increased theta band power in remote areas predicted by functional connectivity, especially when the stimulation was delivered close to white matter \cite{Solomon2018}. These studies show the potential of using direct electrical stimulation in therapeutic neuromodulation, and intracranial stimulation through iEEG can be a useful tool to rapidly explore possible stimulation locations and parameters for the design of effective neuromodulation.

Consistent stimulation effects -electrophysiologically or behaviourally- across sessions are crucial for developing therapeutic neuromodulation treatments. For example, understanding the underlying electrophysiological effect of transcranial stimulation and its consistency is an important step towards taking advantage of its already demonstrated benefits on motor rehabilitation \cite{Pollok2015, Buch2017}. Relevant investigations on cross-session consistency have been reported in non-invasive stimulation modalities (for a review see \cite{Guerra2020}). For instance, while the electrophysiological effect of transcranial magnetic stimulation has been reported to be highly consistent across sessions \cite{Hermsen2016}, while transcranial direct current stimulation (tDCS) effect was found to be inconsistent \cite{Dyke2016, Horvath2016} (but see also \cite{Ammann2017}). The sources of such variability have been discussed extensively in the context of inter-individual studies but some of them apply on an intra-individual basis as well (e.g., baseline physiological state, cognitive task at hand; for a review see \cite{Li2015}). However, to our knowledge, the cross-session consistency of the electrophysiological effects of iEEG stimulation has not yet been systematically investigated.

Here we investigate the consistency of the iEEG stimulation effect in terms of band power modulations between stimulation sessions from the same subject. We measure how stimulation modulates band power in five different frequency bands and investigate whether these modulations vary from one stimulation session to the next for the same subject and stimulation location. We introduce a measure of consistency that accounts for the distributed stimulation effects recorded across multiple iEEG channels. We finally investigate which features of the stimulation protocol, the measured stimulation effect, and the baseline conditions most influence between-session consistency.


\section{Methods}

\subsection{Electrophysiological and cortical surface data}
We used data that are publicly available as part of the Restoring Active Memory (RAM) project (managed by the University of Pennsylvania; \url{http://memory.psych.upenn.edu/RAM}). As stated in the project's website "Informed consent has been obtained from each subject to share their data, and personally identifiable information has been removed to protect subject confidentiality". The original research protocol for data acquisition was approved by the relevant bodies at the participating institutions. Furthermore, the University Ethics Committee at Newcastle University approved the current project involving the data analysis reported here (Ref: 12721/2018). We extracted data from all patients (n=87) that underwent at least one stimulation session while performing memory tasks. We excluded 8 patients that either had substantial stimulation artefacts in almost all channels or their data were limited (single session with <18 stimulation trials). Thus, we analysed data from 79 subjects from which 36 had at least 2 stimulation sessions with the same stimulation location (totalling 101 pairs of sessions with same stimulation location).

\subsection{Stimulation Paradigm}
Stimulation was delivered using charge-balanced biphasic rectangular pulses (300 $\mu$s pulse width) at 10, 25, 50, 100, or 200 Hz frequency 0.25–3.5 mA amplitude. The duration of the stimulation was 500 ms or 4.6 s, depending on the experiment.

\subsection{Preprocessing}

\begin{figure}[H] 
\hspace{-35mm}\includegraphics[width=190mm]{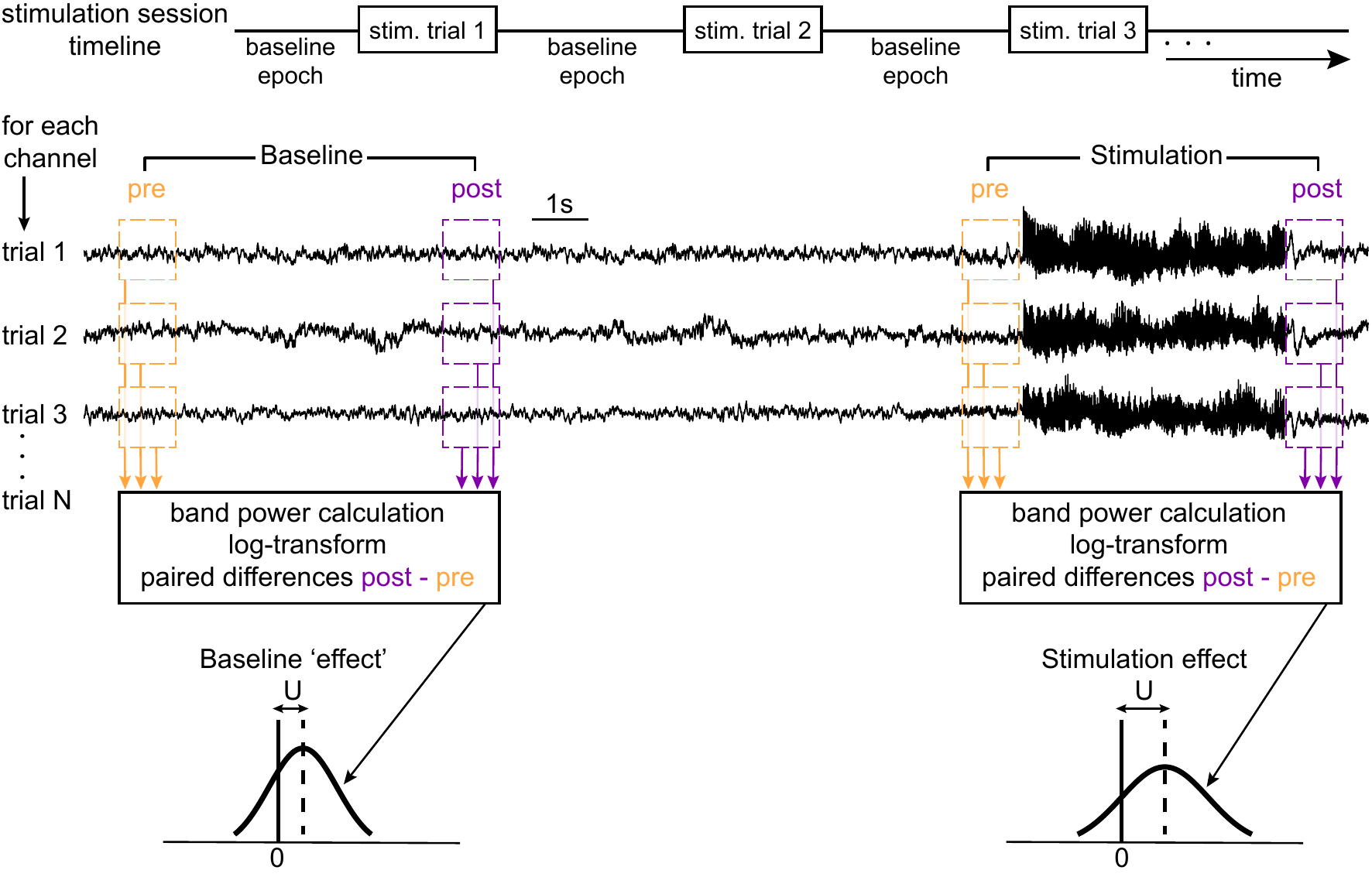}
\caption{{\bf Measuring band power changes in response to stimulation and band power baseline fluctuations.}
Top panel: Timeline of a typical stimulation session. The schematic also shows how pre- and post-stimulation segments are extracted from each stimulation trial and analysed in terms of their band power. While only three trials are shown here, a typical stimulation session had 60 trials (median value with 13.9 SD). Lower panels: Band power in five different frequency bands was calculated and log-transformed for each extracted segment. The effect of stimulation on band power, and equivalently, band power's fluctuations during baseline, are expressed by the effect U, which is derived from a non-parametric test applied to the paired differences between pre and post segments.
}
\label{fig1} 
\end{figure}

To measure stimulation effect, 1-second segments were extracted from the iEEG signals around every stimulation trial; that is, we extracted one segment before (pre) and one after (post) the stimulation event, with a 50ms buffer between each segment and the event. To assess baseline fluctuations, ‘pre’ and ‘post’ segments were also extracted from the baseline activity during baseline epochs, with a pre-post interval equal to the one around the stimulation trials of the same session. A baseline epoch was considered to be any inter-stimulus interval which was at least 20 sec long and 5 sec away from the stimulation itself. Figure \ref{fig1} shows a schematic of the session timeline and the process of segment extraction. Since the stimulation trials were temporally organised in groups of three in a typical session (i.e., less than 10s interval between trials in the same group), we extracted baseline pre/post segments from each baseline epoch in groups of three as well (see Fig. S1 in Supplementary Material), such that the number of segments taken around stimulation and the number of baseline segments were approximately equal in each session. 

The time series of each segment were centred around zero and de-trended. De-trending was achieved by applying linear regression and then removing the least-squares fit from the signal. Any channels with repeated artefacts were excluded (see below). A common average re-referencing was applied to the remaining set of channels. The stimulation channels were excluded from the common average calculation, but the calculated common average was applied to them. The band power of each segment was calculated in 5 different bands [delta (2-4 Hz), theta (4-8 Hz), alpha (8-12 Hz), beta (12-25 Hz), and gamma (25-55 Hz)] after estimating the power spectral density of the segment using Welch’s method (with window length equal to half of the segment length and overlap length equal to a quarter of the segment length). Finally, the band powers were log-transformed. Figure \ref{fig1} shows a schematic of the preprocessing. 

Channels with repeated stimulation artefacts (i.e., voltage deflection) were excluded. A repeated stimulation artefact was detected based on two criteria. Either one of these two criteria was sufficient to indicate a channel with repeated artefacts. First, a strong effect of stimulation on the average (across time) voltage of the first half of the post segment compared to the average (across time) voltage of the second half of the pre segment. The effect was quantified using the t-statistic of a paired t-test. Second, the second half of the average (across trials) post signal had a slow return to the average (across time) voltage value of the pre segments. This was detected by linear regression.


\subsection{Box Plots}
Box plots were used to summarise various distributions in the Results. Central lines indicate  median values, while the boxes extend from the $25^{th}$ to $75^{th}$ percentile (interquartile range) of the distribution. Whiskers extend to the upper and lower adjacent values, that is, the most extreme values that are not outliers. Outliers are considered to be values that lie more than 1.5$\times$[interquartile range] away from the $25^{th}$ or $75^{th}$ percentile.

\subsection{Effect measures}
The effect of stimulation on band power from pre to post was considered as the z-statistic (indicated by U throughout) produced by the Wilcoxon sign rank test (paired non-parametric test; signrank function in MATLAB). A positive U indicates an increase in band power from pre to post, whereas a negative U indicates a decrease from pre to post. To also quantify the baseline fluctuations of band power, the same measure was used on the ‘pre/post’ pairs of the baseline activity (Fig. \ref{fig1}).

The overall difference in stimulation effect between two sessions (across all channel/band combinations) in Fig.~\ref{fig4}B was quantified by using the absolute t-statistic of a paired t-test on the absolute effect U of the two sessions. We used absolute effects as we wanted to generally assess changes in effect magnitude.

\subsection{Consistency coefficient}
The consistency of stimulation effect was measured for each pair of sessions with the same stimulation location in the same subject. All possible combinations of 2 sessions were considered, totalling 101 pairs. The consistency was computed by first pairing the effect values of corresponding channel/band combinations between the two sessions. Note that only the intersection of valid channels between the two sessions was considered (a channel can be excluded due to artefacts in one session but not the other). The consistency coefficient was given by the Fisher-transformed zero-centred Pearson’s correlation. Considering the effect values of the two sessions as random variables $S_1$ and $S_2$, then the consistency coefficient is given by:
$r_0 = \mathbf{E}[S_1S_2]/(\hat{\sigma}_1\hat{\sigma}_2)$,
where $\mathbf{E}$ denotes expected value and $\hat{\sigma}$ refers to the average deviation from 0 ($\hat{\sigma} = \sqrt{(\sum^n_{i=1}{s^2_i})/n}$). We use the zero-centred Pearson’s correlation to only detect a zero-translated agreement between the random variables, that is, in the form of $S_1 = kS_2$, with 0 intercept and $k$ a non-zero constant.

\subsection{Consistency curve}

The consistency curve was used to express the consistency between two sessions by gradually considering fewer pairs of effect values at low effect sizes. Considering a scatter plot of all the effect value pairs, it was computed by gradually increasing the radius of an exclusion circle emanating from (0,0). The consistency curve at radius = 0 gives the consistency when all points are included in the consistency calculation, whereas the consistency curve at radius = x expresses the consistency as computed after excluding every pair of effect values that lie inside a circle with centre (0,0) and radius x. The circle was gradually enlarged with a step of 0.2 and the enlargement stopped just before covering 98\% of the scattered values. We used this procedure to ensure that we can detect consistency even if only a few channels exhibited consistency, without the consistency being masked by low effect channels.

Each consistency curve is represented by its maximum consistency coefficient. The maximum consistency coefficient is the value on the curve that deviates the most from 0, being positive or negative. Thus, it expresses the strongest correlation or anti-correlation found between the effect values of the two sessions.

\subsection{Multiple Linear Regression Analysis}

To explore which factors determine consistency across all 101 session pairs, we modelled the maximal value of consistency as a linear combination of the following variables: 
\begin{itemize}
\item session time difference: absolute time difference between the sessions' starting timestamps. 
\item difference in baseline (band power) means: mean absolute paired difference between the sessions' mean values of band power during baseline (both ‘pre’ and ‘post’).
\item difference in baseline (band power) standard deviations: mean absolute paired difference between the sessions' standard deviations of band power during baseline (both ‘pre’ and ‘post’).
\item average max effect: average (between sessions) maximum effect (across all channel/band combinations).
\item average min effect: average (between sessions) minimum effect (across all channel/band combinations).
\item average stimulation amplitude: average stimulation amplitude between sessions.
\item stimulation amplitude difference: difference in stimulation amplitude between sessions.
\item stimulation frequency: frequency of stimulation pulse train (always common between examined session pairs).
\item depth of the stimulation location: distance of stimulation location (midpoint between anode and cathode) from brain surface.
\item task difference: difference in memory tasks (categorical variable) carried out by the subject during recording; that is, 0 for same and 1 for different tasks between sessions. 
\end{itemize}

The stimulation depth was computed as the Euclidean distance of the anode-cathode midpoint from the subject's brain surface. If that midpoint was found to be outside the provided surface, its depth was set to negative (minus the Euclidean distance).

\subsection{ANOVA test}
In order to quantify the explanatory power of all the different independent variables on the consistency we used ANOVA test on the model built by the Multiple Linear Regression Analysis. We built the model and assessed the ANOVA effects 200 times through bootstrapping.
We used this bootstrapping approach to check for the robustness of the model. The ANOVA effect, the $R^2$, and the Adjusted $R^2$ are reported.

\section{Results}

\subsection{Stimulation elicits a weak effect in most sessions and across frequency bands}

\begin{figure}[H] 
\hspace{-30mm}\includegraphics{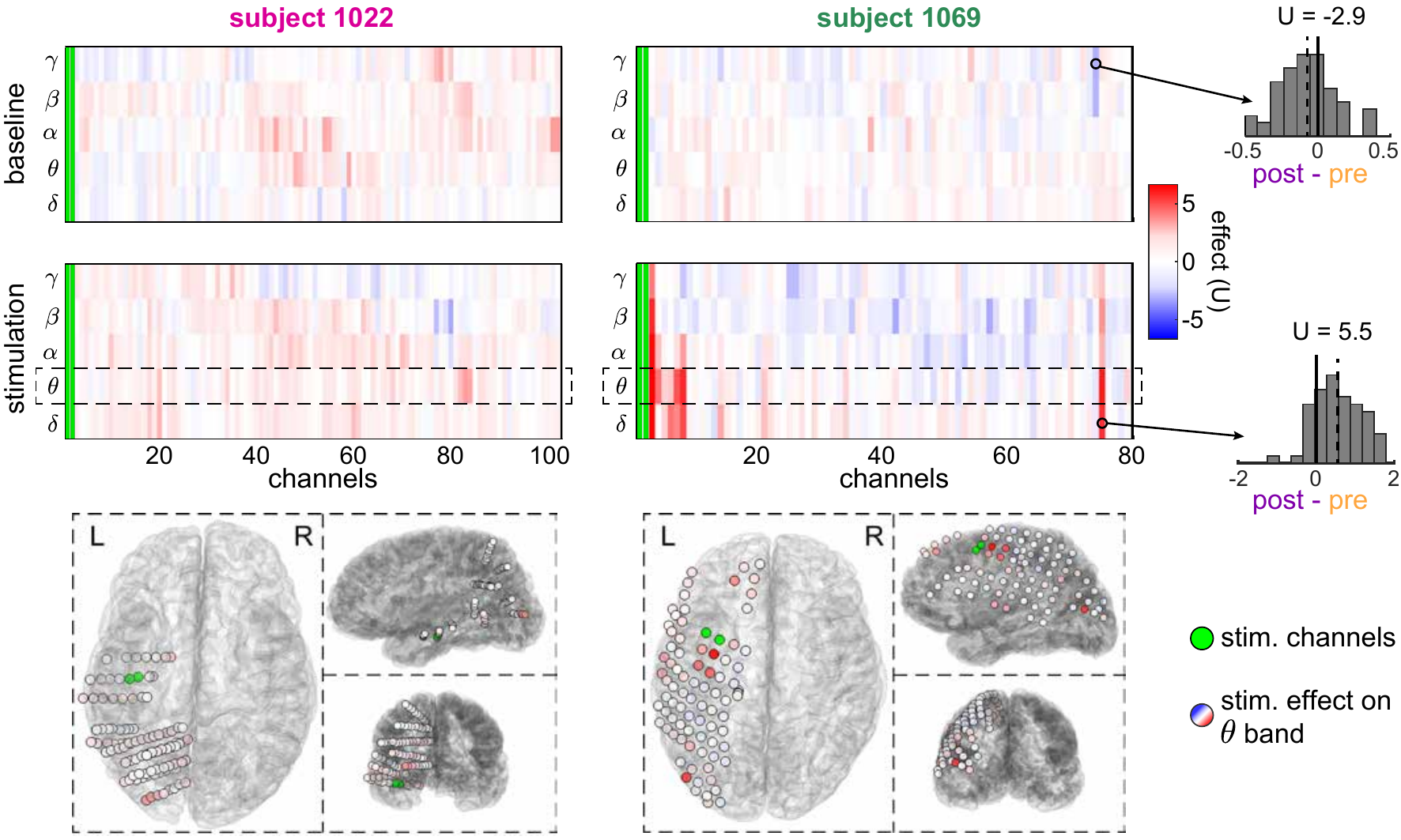}
\caption{{\bf Examples of sessions with low and high stimulation effect.}
The heat maps show the stimulation effect in two example sessions: one from subject 1022 with low effect and one from subject 1069 with high effect. The effect was measured for all combinations of channels and frequency bands. Notice that the effect can be positive or negative, indicating increase or decrease of band power from pre to post stimulation (see example distributions of the differences (post-pre) in the rightmost panels). The fluctuations of band power during baseline are also shown for comparison. The channels are sorted based on their Euclidean distance from the stimulation site. The lower panels show the spatial distribution of the stimulation effect on theta band across the cortex.}
\label{fig2} 
\end{figure}

Figure \ref{fig2} shows the measured stimulation effects across channels and frequency bands for one example session in each of two example subjects 1022 and 1069. These example sessions represent sessions with weak (Fig. \ref{fig2}, left) and strong (Fig. \ref{fig2}, right) stimulation effects. As a reference, the upper panels show the ``effect'' during baseline, that is, the background fluctuations of band power. The lower panels show the stimulation effect in terms of band power changes, based on multiple pre- and post-stimulation pairs (see example inset panels on the right and Fig. \ref{fig1}). Notice that, even in the example subject 1069, where some strong stimulation effects are seen, these are restricted to a handful of channels and specific frequency bands. This observation is typical for all the sessions that exhibited a strong effect. Similarly, the example session on the left is a typical example of all the sessions that have a stimulation effect that is indistinguishable from the baseline fluctuations.

The lower panels in Fig. \ref{fig2} show the spatial layout of the iEEG stimulation and recording channels in the brain, with electrodes colour-coded by their corresponding stimulation effect sizes. Note that a strong stimulation effect, in this case on theta band, is not limited to contacts close to the stimulation site but also affected remote contacts (lower right panel).

\begin{figure}[H] 
\centering\includegraphics{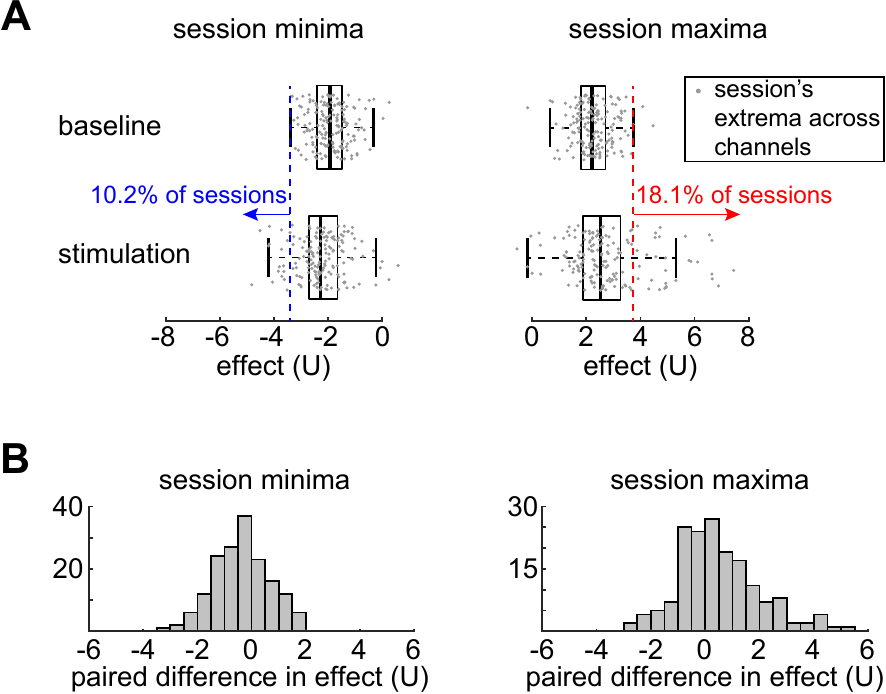}
\caption{{\bf Low stimulation effect on theta band found in most sessions.} \textbf{A} 
Across all sessions, the distributions of their extrema effect values on theta band are compared between baseline and stimulation. Each point corresponds to the minimum (left) or maximum (right) effect value U of all recording channels in a given session. A minority of sessions have stimulation extrema (10.2\% for min and 18.1\% for max) that are more extreme than the adjacent values seen in baseline distributions (adjacent values being the most extreme values that are not outliers). \textbf{B} The histograms present the paired (per session) differences in extreme values of effect on theta band (session stimulation effect – session baseline effect).}
\label{fig3} 
\end{figure}

In order to assess if the effect of stimulation exceeded baseline fluctuations in general across all 165 sessions and 79 patients, we compared the extrema of the stimulation effect to the extrema of the baseline fluctuations for each frequency band. Figure \ref{fig3}A shows the distributions of minima and maxima effect on theta band for baseline and stimulation. These extrema were taken across channels to capture the strongest effect during a session. Generally, it is evident that, even the channel with the strongest stimulation effect does not have a substantially larger effect size compared to the baseline fluctuations. In theta band, only 10.2\% of the sessions exhibit a minimum (negative) stimulation effect that exceeds the adjacent value of the baseline minima. Similarly, only 18.1\% of the sessions exhibit a maximum (positive) stimulation effect that exceeds the adjacent value of the baseline maxima (see Fig. \ref{fig3}A). 

The limited stimulation effect across all sessions was also evident when we computed the paired differences in effect between stimulation and baseline conditions. The histograms for the effect minima and maxima in Figure
\ref{fig3}B indicate that, in most sessions, even the most extreme effect sizes do not exceed the band power fluctuations during
baseline. However, these distributions are not zero-centred (paired t-test for minima: $p = 2.4 \cdot 10^{-5}$, effect size for minima: -0.430; paired t-test for maxima : $p = 4.9 \cdot 10^{-5}$, effect size for maxima: 0.437), indicating that across patients and sessions there is a small but significant difference between baseline and stimulation conditions in our dataset. Similar results were found for all frequency bands (see Fig. S2 in Supplementary Material).

\subsection{Limited effect of stimulation is not due to low stimulation amplitude}

\begin{figure}[H] 
\centering\includegraphics{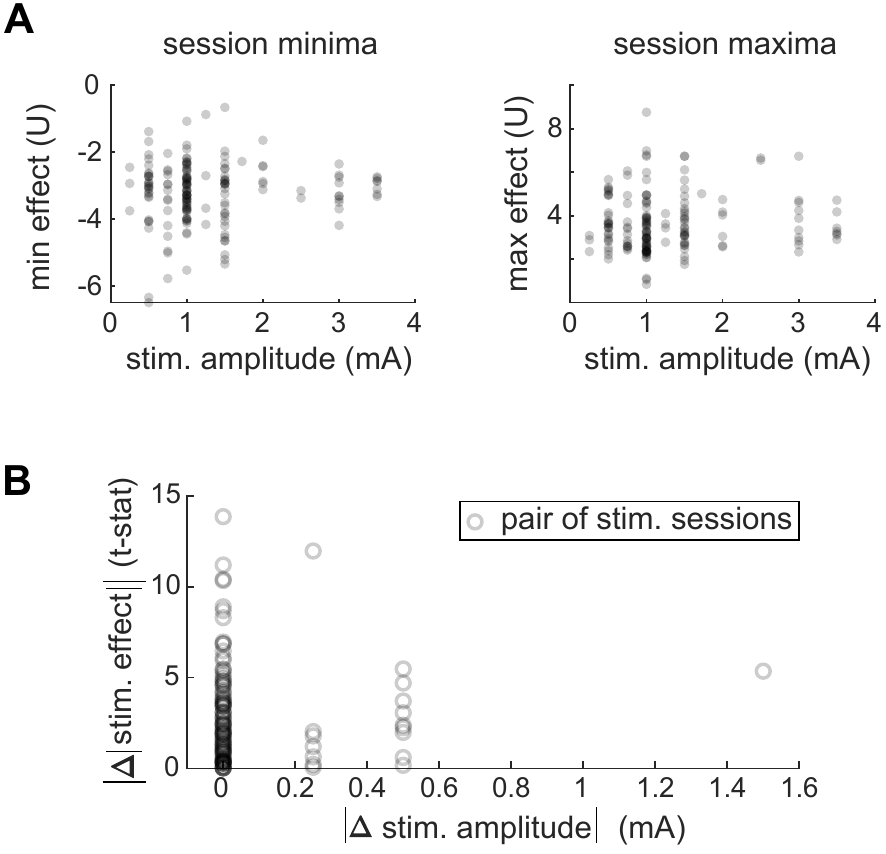}
\caption{{\bf No correlation found between stimulation amplitude and effect.}
\textbf{A} The effect minima and maxima (across all channels and frequency bands) from each session is scattered against the stimulation amplitude of the session. 
\textbf{B} 
For all the pairs of sessions that come from the same subject and have the same stimulation location, the difference in their stimulation effects is scattered versus the difference in their stimulation amplitudes. 
}
\label{fig4} 
\end{figure}

Next, we investigated whether the low stimulation effect size in most sessions can be attributed to the stimulation amplitude of the session. Figure \ref{fig4}A shows that there is no correlation between the effect size achieved in the session and the session stimulation amplitude. The distributions of effect sizes for each session, across all channels and frequency bands, are represented by their minima and maxima. Neither of these two measures tend to increase or decrease with the stimulation amplitude (range: 0.25 - 3.5 mA; see also Fig. S3 in Supplementary Material for band specific results).

Furthermore, we considered all the pairs of stimulation sessions with the same stimulation location in the same subject (101 pairs). We tested whether their difference in effect size is correlated with the difference of stimulation amplitude between the sessions. Figure \ref{fig4}B shows that the absolute difference in effect size is not correlated with the absolute difference in stimulation amplitude. Thus, even for the same subject and the same stimulation location, an increase in stimulation amplitude does not necessarily produce a stronger effect.

\subsection{Weak stimulation effects are inconsistent across sessions}

\begin{figure}[H] 
\hspace{-5mm}\includegraphics{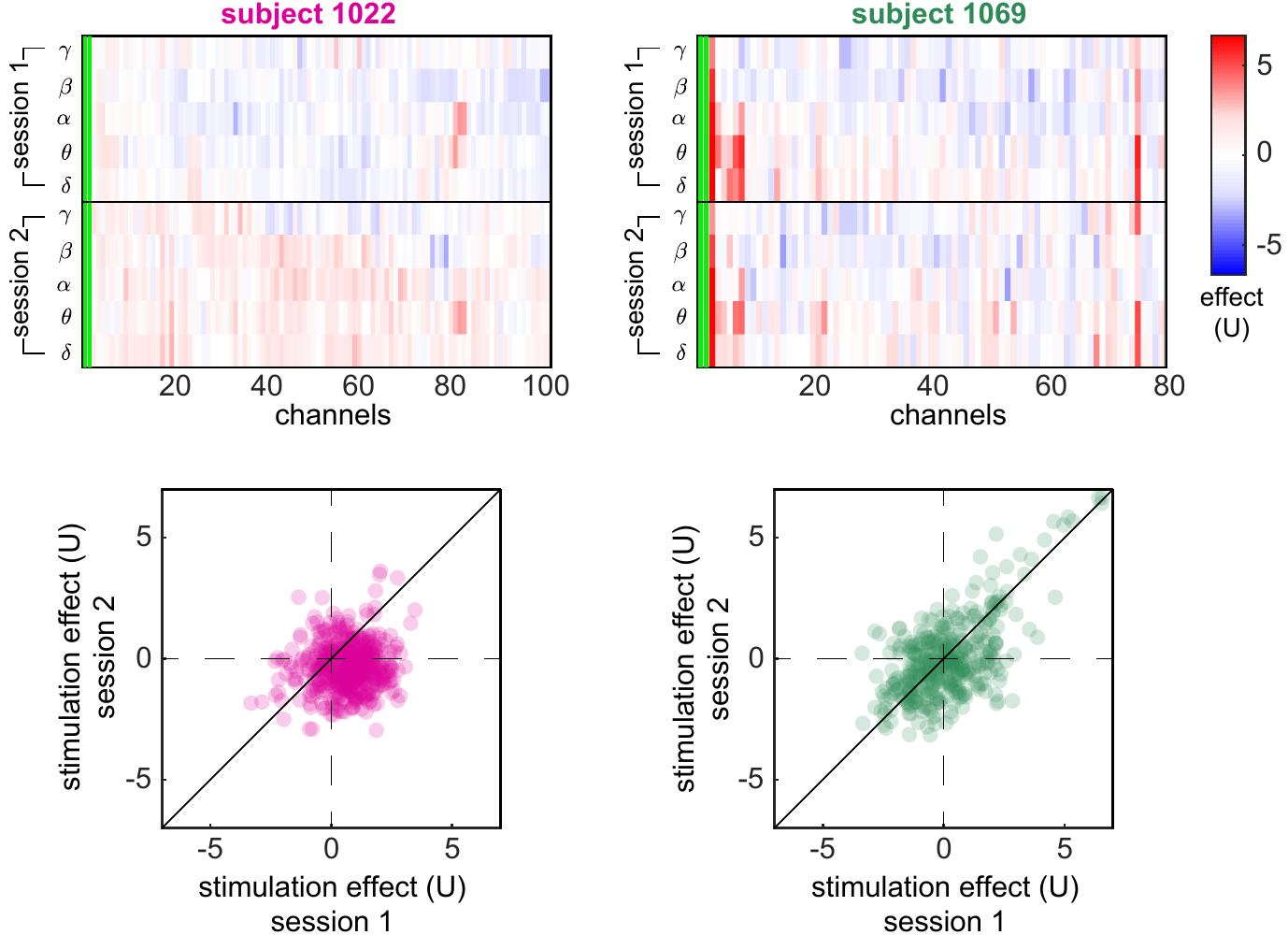}
\caption{{\bf Examples of session pairs with low and high effect consistency.}
The stimulation effect in two pairs of sessions, from two
different subjects, is shown in the top panels as examples. The pair of sessions on the left (subject 1022) has low consistency whereas the pair on the right (subject 1069) has high consistency. This disparity is more clearly shown in the lower panels, where the corresponding effect pairs for each channel and frequency band are scattered. The low consistency on the left is shown by a circular cloud of points, whereas the high consistency on the right is shown by an elongated cloud of points.}
\label{fig5} 
\end{figure}

To investigate the consistency of stimulation effect across sessions, we focused on pairs of sessions in the same subject and stimulation location. Figure \ref{fig5} shows two examples of these session pairs. The example on the left (subject 1022) does not show positive correlation between the two sessions in terms of stimulation effect in different channels and frequency bands. The example on the right (subject 1069) shows that the patient’s two sessions are positively correlated. Notice that this correlation is mainly driven by channels that exhibit a strong positive stimulation effect in the first place.

\begin{figure}[H] 
\vspace{-25mm}\hspace{-10mm}\includegraphics{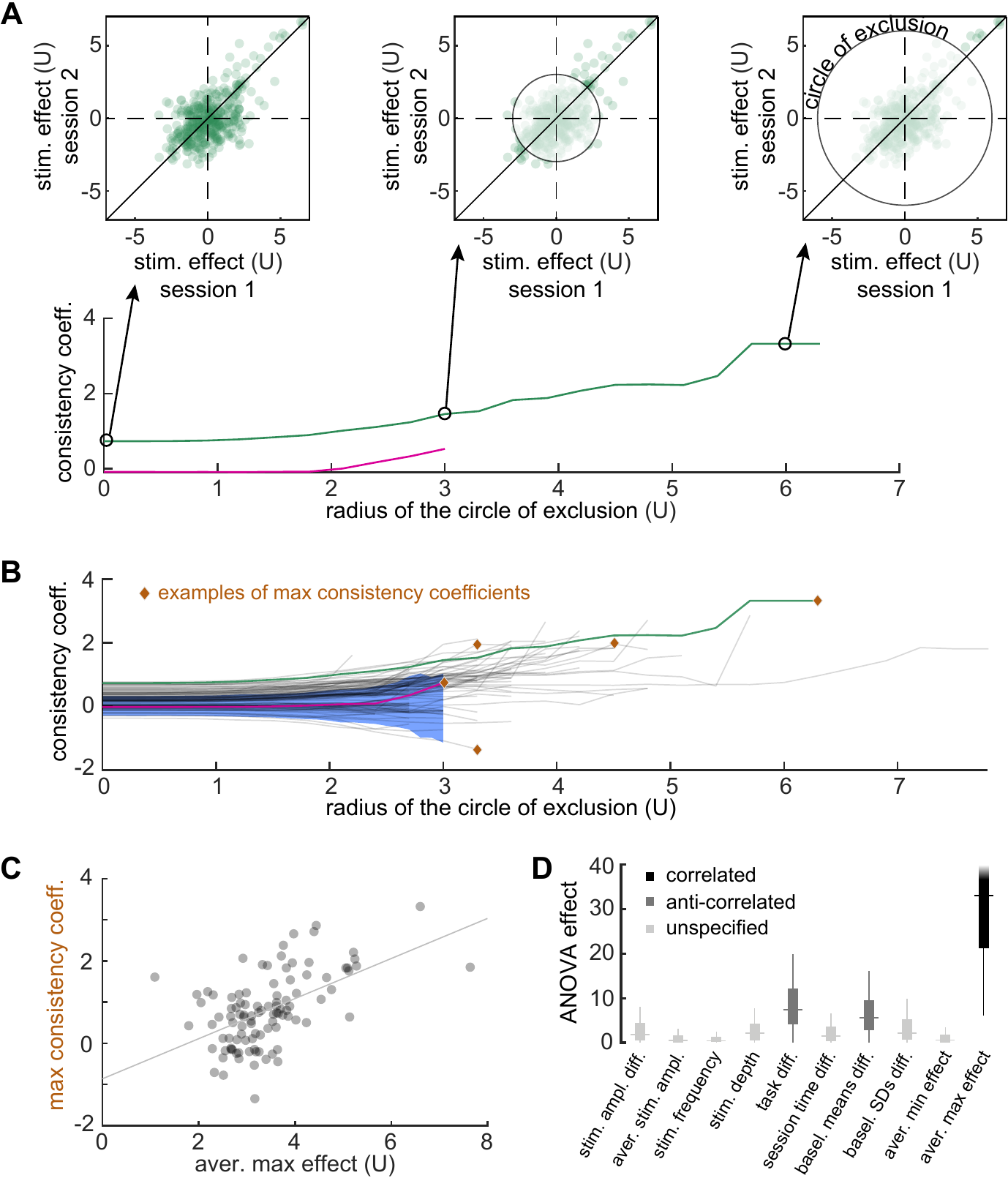}
\caption{{\bf Cross-session consistency is found in a minority of subjects while it relies heavily on strong (positive) effect.}
\textbf{A} Consistency curves were computed by gradually enlarging the circle of exclusion and calculating the consistency coefficient on the remaining scatter points. Three example radii for the circle of exclusion are shown. The plotted consistency curves represent the two example session pairs in Fig. \ref{fig5}.
\textbf{B} All 101 consistency curves, one for each session pair, are shown with five examples of maximal consistency coefficients achieved (brown diamonds). The shaded blue region indicates the 95\% two-sided confidence interval of the consistency coefficients of baseline activity.
\textbf{C} Maximum consistency coefficient scattered versus average maximum effect reveals a strong correlation between them (Pearson's $r=0.536, p=7.4\times10^{-9}$).
\textbf{D} Distributions of ANOVA effect values (produced through bootstrapping - see Methods) for the independent variables used in the multiple linear regression model which was used to explain the maximum consistency coefficients. The average maximum effect between paired sessions has the strongest explanatory power over consistency. Both task difference and difference in baseline mean have a fair explanatory power over consistency. Outliers are omitted for clarity.
}
\label{fig6} 
\end{figure}

In order to assess the level of consistency in stimulation effect across all 101 session pairs, we computed the consistency curve for each pair. Figure \ref{fig6}A shows the consistency curves of the two session pair examples in Fig. \ref{fig5} alongside some illustrations on how the curve is computed: a circle of exclusion emanating from (0,0) is gradually enlarged and the consistency coefficient is calculated for varying values of the circle’s radius (see Methods). The consistency curve (as a function of the radius) captures the consistency coefficient of the session pair when all effect values are considered (at radius 0) but also while increasingly excluding channel and frequency band combinations with weaker stimulation effect (at higher radii). The exclusion of channel/band combinations with weak effect serves to minimise the influence of the inherently inconsistent band power fluctuations on the consistency calculation. In addition, considering the selective connectivity of brain areas, it is expected that only a subset of channels will respond to a localised stimulation. The two curves shown in Figure \ref{fig6}A capture the difference between high and low consistency as shown in the scatter plots of Fig. \ref{fig5}, not only when all values are considered, but also when only the strong effect values are considered. This approach of gradually excluding the weaker stimulation effects (around the level of baseline fluctuations) essentially allows us to capture consistency in the few channels that display a discernible stimulation effect in the first place.

Figure \ref{fig6}B shows the consistency curves for all 101 session pairs and the confidence interval of consistency coefficients of the baseline periods (blue background). The overall consistency in this dataset is not high: 32.7\% of the session pairs have higher consistency than the $97.5^{th}$ percentile of the baseline `effect’ at radius = 0; 12.9\% of the session pairs have higher consistency than the $97.5^{th}$ percentile of the baseline `effect’ at radius = 3; and only 34.6\% of the session pairs have a maximum consistency that is higher than the maximum value of the baseline `effect’ confidence interval. Four examples of maximum consistency coefficients on four of these curves are indicated with brown markers in Fig. \ref{fig6}B. We will consider these maximum consistency coefficients as a representative value of the session pair consistency in the following (i.e., highest consistency achieved after exclusion of some not stimulation-related channels).

In Fig. \ref{fig6}C, we demonstrate a strong and significant correlation between the average maximum effect and the maximum consistency coefficients (Pearson's $r=0.536, p=7.4\times10^{-9}$). Theta and alpha bands contribute more to this correlation  (see Fig. S4 in Supplementary Material). As a comparison, we applied the same procedure to simulated data (normal distribution with mean 0 and standard deviation matching the the sessions’ baseline), and the correlation is not present (Pearson's $r= -0.003, p=0.438$). Essentially,  the stronger stimulation effects also tend to be more consistent across sessions. 

Finally, we built a multiple linear regression model to explain the maximum consistency coefficients as a linear combination of multiple independent variables including the average maximum effect ($R^2 = 0.406$, Adjusted $R^2 =0.340$). The high explanatory power of the average maximum effect is also evident after running ANOVA on the multiple linear regression model, with the results shown in Fig. \ref{fig6}D (distributions produced after 200 bootstrap samples). Other than the strong effect of the average maximum effect on consistency, Fig. \ref{fig6}D shows a fair effect of both the task difference and the difference of baseline mean on consistency ($p=0.008$ and $p=0.017$, respectively), which are both anti-correlated with the maximum consistency coefficient.

\section{Discussion}

We showed that the cross-session consistency of stimulation effect (in terms of band power modulations) is relatively low in a group of 36 subjects who had multiple stimulation sessions through iEEG. A third of session pairs indicate a consistency that is above the baseline consistency (Fig. \ref{fig6}B). High consistency of stimulation effect was found to rely heavily on a strong positive effect of stimulation, that is, high increase of band power (Fig. \ref{fig6}D). Thus, given these findings, the low consistency levels would be expected in this dataset since the stimulation effect was limited (Fig. \ref{fig3}). Other datasets with more pronounced stimulation effect in terms of band power changes may exhibit a higher level of consistency between sessions.

Variability in the baseline brain state may have impacted the consistency of the stimulation responses in our data set. Even the stimulation response within a session has been repeatedly found to depend on the underlying brain state \cite{Papadopoulos2020, Ewell2015, Lesser2008}. This finding is corroborated here since consistency was found to be anti-correlated with both the difference in baseline mean band power and difference in memory task which can be understood as a difference in brain state (Fig. \ref{fig6}D). In other words, the more similar the brain states (as measured by task, or band power configurations) were in this dataset, the more consistent the stimulation effects tended to be. Therefore, a practical advice is to use the same task across stimulation sessions if consistency across sessions is desired.

Our multiple linear regression model included the stimulation depth as one of the independent variables, and it did not exhibit a strong predictive power over consistency. This is not surprising since we did not find any strong relation between stimulation depth and the effect U in the first place (see Fig. S5 in Supplementary Material). However, it is worth noting that there was no distinction between stimulation through surface and depth electrodes in our analysis. The difference between these two types of electrodes cannot be fully captured by the stimulation depth variable. Other confounding characteristics, like the physical dimensions of the contacts and the average distance from other recording electrodes, were not accounted for. Future work can investigate further whether consistency depends on such factors.

Considering the data across all subjects, the most represented stimulation site is the right medial temporal lobe, but several other areas were stimulated. In addition, the spatial extent of the recording electrodes across the dataset covers the whole cortex. A visual inspection of the stimulation sites and the highly responsive sites did not reveal any specific area that was associated with high effect or consistency (see Fig. S6 in Supplementary Material).

Surprisingly, the effect on band power was not correlated with the amplitude of stimulation in this dataset. This finding agrees with the reported insensitivity of motor-cortical excitability to tDCS intensity increases \cite{Jamil2017}. However, another iEEG study has found stimulation intensity to correlate with high frequency activity (30-100Hz), a frequency range which extends beyond those we investigated \cite{Mohan2019}. Furthermore, multiple studies have reported correlations between stimulation intensity and motor improvements when deep brain stimulation of subthalamic nuclei is used for the treatment of Parkinson’s disease (e.g., \cite{Deli2011}). This discrepancy might indicate a non-trivial or non-linear relationship between the electrophysiological and behavioural effects of an increasing stimulation intensity. The potentially `all-or-nothing' response may further depend on the stimulated area.

In our study, the stimulation effect was measured based on the immediate responses within a session only. Arguably, the effect of stimulation can manifest at longer timescales or in other features and those effects may be more consistent across sessions \cite{Huang2005, Suppa2016}. This also relates to our definition of baseline in this study. Segments of baseline are taken from interstimulus intervals that may carry some post-stimulus modulations of band power. Any consistency in long-term changes due to stimulation should be investigated in future studies.

Cross-session consistency of stimulation effect is critical for developing therapeutic neuromodulation treatments, both in terms of electrophysiological, as well as behavioural stimulation effect. This is supported by recent studies which established relationships between stimulation-induced modulation of specific frequency bands and behavioural outcomes \cite{ Muller2018, Natu2019}. Despite the fact that some anatomical factors (e.g., thicknesses of the skull and the cerebrospinal fluid layer) do not influence intracranial stimulation, as opposed to tDCS \cite{Opitz2015}, we found that stimulation through iEEG still has low consistency in terms of band power modulations across sessions in our dataset, similar to tDCS \cite{Dyke2016, Horvath2016, Ammann2017}. Our results suggest that ensuring a strong positive modulation of band power through stimulation, by choosing the appropriate stimulation location and parameters, is prerequisite for a high consistency across sessions. In addition, our results suggest that the dynamical brain state needs to be taken into account and a state-depended framework of stimulation may be required. The present and previous studies all show that more sophisticated protocol designs are needed to maximise the benefit of neurostimulation interventions.

\section*{Acknowledgments}
We thank the CNNP team (www.cnnp-lab.com) for discussions on the manuscript and the presentation of the results. 
CAP, PNT, and YW gratefully acknowledge funding from Wellcome Trust (208940/Z/17/Z and 210109/Z/18/Z).

\bibliographystyle{unsrt} 
\bibliography{myBib.bib} 

\begin{itemize}
\item
{\textbf{Supplementary Material}\\
Set of supplementary figures}
\label{SupplMat} 
\end{itemize}

\end{document}